# On the Availability of ESO Data Papers on arXiv/astro-ph

Uta Grothkopf[1]
Dominic Bordelon[1]
Silvia Meakins[1]
Eric Emsellem[1]

[1] ESO

Using the ESO Telescope Bibliography database telbib, we have investigated the percentage of ESO data papers that were submitted to the arXiv/astro-ph e-print server and that are therefore free to read. Our study revealed an availability of up to 96 % of telbib papers on arXiv over the years 2010 to 2017. We also compared the citation counts of arXiv *vs.* non-arXiv papers and found that on average, papers submitted to arXiv are cited 2.8 times more often than those not on arXiv. While simulations suggest that these findings are statistically significant, we cannot yet draw firm conclusions as to the main cause of these differences.

The ESO Telescope Bibliography telbib is a database of refereed papers published by the ESO user community[1]; all telbib data papers use at least some data from ESO facilities. The telbib database is curated and further developed by the ESO librarians. Records in the database are linked to the ESO Science Archive, enabling easy access from the literature to the data used in the papers and *vice versa*. In addition, telbib entries are comprehensively tagged and annotated, allowing us to derive visualisations, statistics and reports for various parameters. The telbib content spans more than 20 years, beginning with the publication year 1996, and the database has become an essential tool in understanding publishing trends among the ESO user community. Further information about telbib is available on the web[2,3].

As of July 2017, telbib included almost 13 600 records. 95 % of these papers were published in the core astronomy journals, namely in Astronomy & Astrophysics (A&A; 49 %), the former Astronomy & Astrophysics Supplement Series (A&AS; 2 %), the Astrophysical Journal (ApJ; 20 %), the Astrophysical Journal Supplement Series (ApJS; 1 %), the Monthly Notices of the Royal Astronomical Society (MNRAS; 19 %), and the Astronomical Journal (AJ; 4 %). A more detailed distribution of ESO data papers across journals can be seen in Figure 1.

The ESO Library provides journal access to ESO staff through subscriptions. In astronomy, the most prominent journals (A&A, ApJ/ApJS, AJ and MNRAS) are community owned and operated. This enables astronomers to influence developments in journal publishing to a large extent, including changes to subscription fees and page charges. This differs across scientific disciplines. Steep increases in subscription fees to scholarly journals, along with widespread public access to the internet, led to the start of the Open Access (OA) movement in the 1990s.

## Open Access publications

According to the Scholarly Publishing and Academic Resources Coalition (SPARC), Open Access in scholarly literature is defined to be the "free, immediate, online availability of research articles coupled with the rights to use these articles fully in the digital environment"[4]. This means that users can read, download, copy, distribute, and text mine articles, or use them for any other lawful purpose. Creative Commons copyright licenses define standardised ways of sharing and reusing. The most commonly-used license, called "CC BY Attribution", allows a wide range of reuses provided that credit is given to the original creators. An overview of CC licenses can be found on the Creative Commons website[5].

Different classifications exist to describe the various flavours of open access; the most common versions are as follows:
1. Green OA: Papers are typically published in subscription-based journals, but authors self-archive manuscripts on their webpages or submit them to institutional or subject-based repositories (for example, arXiv/astro-ph) where the articles are free to read for everyone. Green OA does not require a CC license.
2. Gold OA: Articles are published in OA journals. Authors, institutions, or other stakeholders pay a publication fee so that the articles become available to everyone for all permitted uses. These are governed by CC licenses. Some publishers offer the option for authors to pay an article processing charge (APC) to make individual articles available through open access in otherwise subscription-based journals; this is then called "hybrid OA".
3. Delayed and temporary OA: Responding to the mandates issued by governments and funding agencies, many journals are applying delayed OA by providing free access to their publications after a given time. Most core astronomy journals are currently available and are free to read after one year, with the exception of MNRAS, which is only free to read three years after publication. Since these articles were originally published in a subscription-based (toll-access) journal, there is no CC license. Alternatively, publishers may choose to provide free access to specific sections or articles — for example, the latest issue — for a specific amount of time.

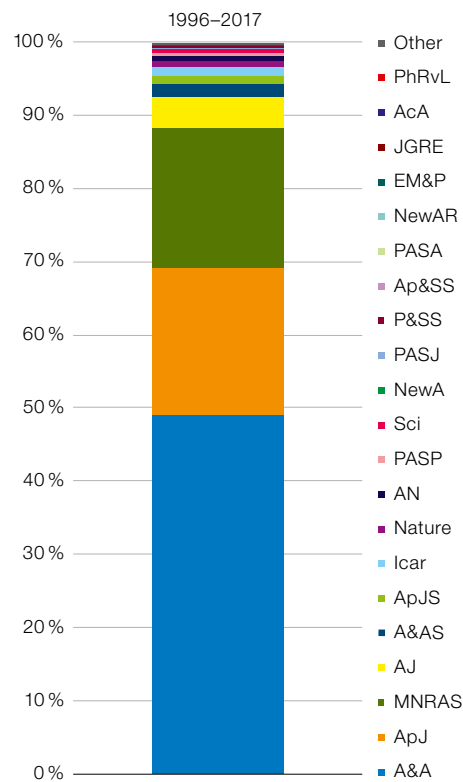

Figure 1. Journal distribution of ESO data papers published between January 1996 and July 2017 (total number: 13 569). Journals in which less than three telbib papers were published are categorised as Other.

In a recent article, Piwowar et al. (2017) undertook a large-scale analysis of the dissemination of OA articles in the sciences. They found that at least 28% of the scholarly literature is OA, equivalent to 19 million articles out of the 67 million included in the study. By far the largest number falls in a category called "Bronze OA" by the authors, meaning that the articles are free to read on the publisher's website but do not have an explicit open access license (similar to the delayed/temporary OA mentioned above). In astronomy, the huge archive of historic literature pertains to this category, including core journals such as A&A, ApJ, AJ and MNRAS spanning back to the first volumes. These historic volumes were scanned by the National Aeronautics and Space Administration Astrophysics Data System (NASA ADS) Abstract Service in collaboration with the Wolbach Library at the Harvard-Smithsonian Center for Astrophysics, and are freely available to the community (Eichhorn et al., 2003).

Piwowar et al. (2017) also studied the so-called OA citation advantage in order to verify whether open access articles are actually cited more often. The authors report that on average, OA articles receive 18% more citations. In this paper, we study the fraction of papers published between January 2010 and July 2017 that are included in the telbib database and have been submitted to arXiv/astro-ph. We also investigate whether there is any difference in impact (as measured by the number of citations) between papers that were submitted to arXiv and those that were not.

### telbib papers and arXiv/astro-ph

The astronomy community has a strong culture of sharing research products, including scientific papers, observatory publications, data, presentation slides, software, code, etc. It is not surprising that astronomers were among the early adopters of the arXiv/astro-ph e-print server and continue to be one of the strongest user groups today, according to arXiv submission rate statistics[6].

| Number of papers | 2010 | 2011 | 2012 | 2013 | 2014 | 2015 | 2016 | 2017 |
|---|---|---|---|---|---|---|---|---|
| All telbib | 738 | 786 | 866 | 842 | 871 | 865 | 944 | 601 |
| Posted on arXiv | 653 | 712 | 793 | 787 | 822 | 827 | 909 | 567 |
| Non-arXiv | 85 | 74 | 73 | 55 | 49 | 38 | 35 | 34 |
| Percentage arXiv | 88.5% | 90.6% | 91.6% | 93.5% | 94.4% | 95.6% | 96.3% | 94.3% |

Table 1. Number of papers per year per category: all telbib papers, arXiv papers, non-arXiv papers and the percentage of telbib papers posted on arXiv.

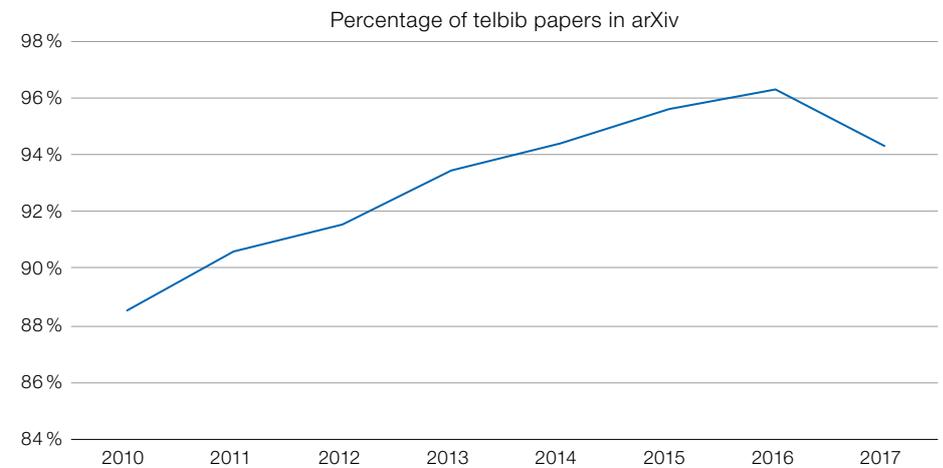

Figure 2. Percentage of telbib papers submitted to arXiv between January 2010 and July 2017.

Taking this tradition into account, we investigate the extent to which papers that use ESO data are free to read. We do not include all flavours of OA, but merely study whether or not papers included in telbib have been submitted to arXiv. Articles that are available otherwise, be it through open access journals, via delayed OA, or as part of a section made available free of charge for a given time by the publisher, are not defined as free to read here unless they were also posted on arXiv.

### Methodology

We used telbib to obtain a list of NASA ADS bibliographic reference codes (bibcodes) for ESO data papers for each year. This list was posted to the ADS Search application programming interface (API) requesting the bibcode, identifier and citation count fields. The identifier field is an array of identifiers that ADS stores, including bibcodes, DOIs and arXiv IDs.

In the result set, we searched each record's list of identifiers for either an arXiv bibcode (i.e., a bibcode created for an ADS preprint and replaced by a journal-based bibcode after publication) or an arXiv identifier with its own consistent pattern. We catalogued these as arXiv papers accordingly.

After querying the ADS API, we took an additional verification step of querying arXiv for the papers that seemed to be "non-arXiv". We searched by author, year and title fragments, then manually compared result records to their ADS counterparts. Using this method we found approximately 30 additional papers that were hosted on arXiv across the seven-year span. These papers were subsequently assigned to the arXiv paper category.

### Fraction of ESO data papers on arXiv/astro-ph

Our study focused on refereed papers published from January 2010 to July 2017 in the ESO Telescope Bibliography. In 2010, 88.5% of the telbib papers were submitted to arXiv. A steady increase can be seen over the following years until an impressive submission rate of 96.3% is reached in 2016. The percentage seems to be lower in 2017, but only the first seven months of this year could be considered.

| Total number of citations | 2010 | 2011 | 2012 | 2013 | 2014 | 2015 | 2016 | 2017 |
|---|---|---|---|---|---|---|---|---|
| All telbib | 33.776 | 29.693 | 27.783 | 26.453 | 20.046 | 14.630 | 8.226 | 1.953 |
| Posted on arXiv | 32.251 | 28.409 | 26.516 | 25.909 | 19.534 | 14.408 | 8.138 | 1.939 |
| Non-arXiv | 1.525 | 1.284 | 1.267 | 544 | 512 | 222 | 88 | 14 |

| Median number of citations | | | | | | | | |
|---|---|---|---|---|---|---|---|---|
| All telbib | 26 | 24 | 21 | 19 | 15 | 11 | 5 | 2 |
| Posted on arXiv | 30 | 26 | 22 | 21 | 16 | 11 | 6 | 2 |
| Non-arXiv | 10 | 12.5 | 11 | 7 | 6 | 3 | 2 | 0 |
| Ratio arXiv *vs.* non-arXiv | 3.0 | 2.1 | 2.0 | 3.0 | 2.7 | 3.7 | 3.0 | |

Table 2. Total and median number of citations: all telbib papers, papers posted on arXiv, and papers not posted on arXiv, along with the ratio of arXiv *vs.* non-arXiv median citations (from January 2010 to July 2017; citations as of 28 August 2017).

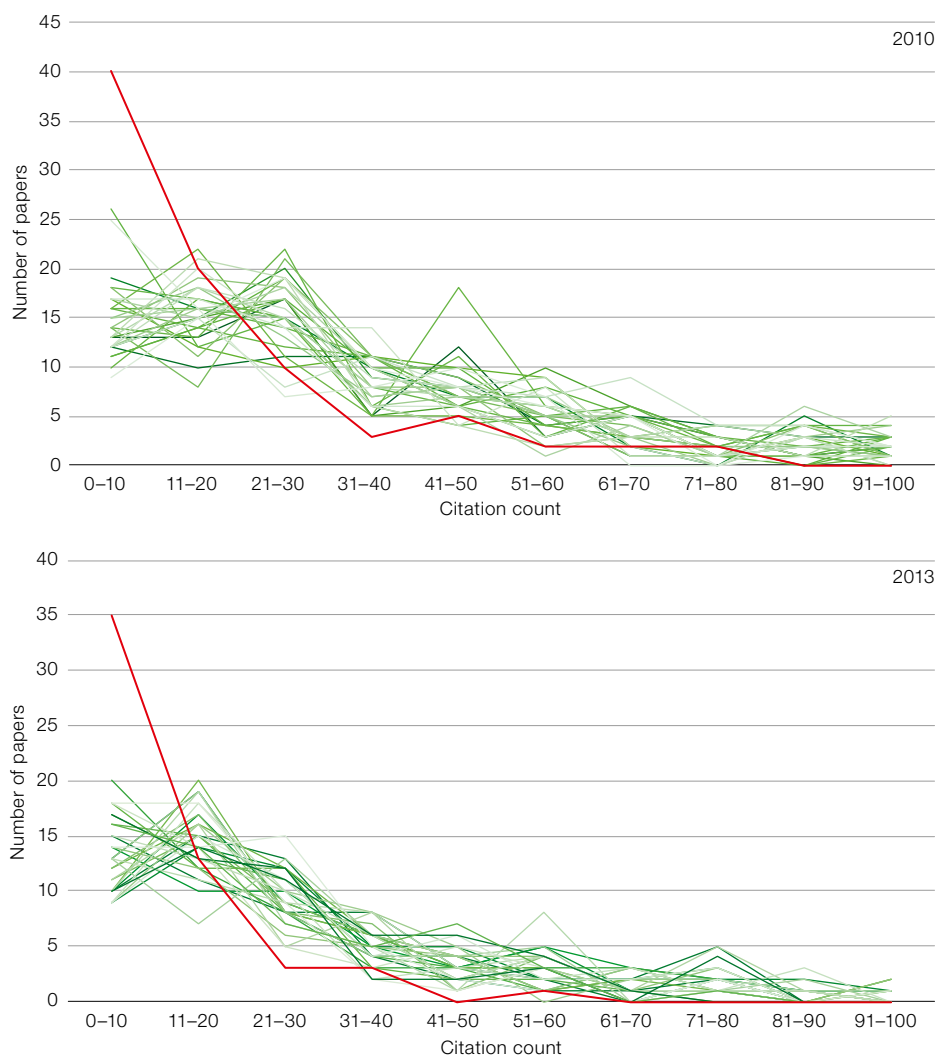

Figure 3. The distribution of citations for non-arXiv papers published in 2010 and 2013 (red line) *vs.* realisations of the same number of randomly selected papers from the entire set of telbib papers published in that year (green lines) for 2010 (top: 85 papers) and 2013 (bottom: 55 papers).

The detailed results are given in Table 1, while Figure 2 shows the percentage of telbib papers in arXiv by year.

Looking in more detail at the non-arXiv papers, we found a wide variety of author affiliations and ESO facilities that provided the observations for these papers. In addition, non-arXiv manuscripts were published in all the major astronomy journals. We also investigated the impact of these non-arXiv papers, where impact is defined as the number of citations relative to the median AJ citation (Bordelon et al., 2016). The most cited non-arXiv papers have an impact of approximately 7–10 for the publication years 2010–2016, with one paper reaching an impact of greater than 17. This means that these specific papers have been cited considerably more often than the median AJ paper of the same year. However, such high-impact papers appear to be outliers. In summary, no obvious pattern can be identified at first glance that would explain why some papers were not submitted to arXiv.

In order to further investigate possible reasons for not submitting a paper to arXiv, we conducted an (unrepresentative) survey among some of the authors of non-arXiv papers (~ 20 % of the 2016 and 2017 papers), asking them about their motivation. While we only received a response rate of about 60 %, we found that there are various reasons for non-submission. Most frequently, authors were too busy for the submission process and had other priorities. One author stated that arXiv was not considered essential for the publishing process.

### Differences in citation counts of ESO data papers posted on arXiv *vs.* non-arXiv

In a second step, we investigated whether arXiv papers are indeed cited more often. For the same set of papers as before (telbib papers published between January 2010 and July 2017), we obtained the citations from the ADS and calculated the total and median citations for the two groups (Table 2).

Comparing the median citations for the years 2010–2016, we found that manuscripts submitted to arXiv received on average 2.8 times more citations than the non-arXiv papers, despite the few high-impact non-arXiv papers mentioned above. Naturally, the most recent papers have not gathered enough citations to achieve a meaningful result. Kolmogorov-Smirnov tests have been conducted to verify our findings. The results suggest that the differences between arXiv and non-arXiv papers are very significant.

We also compared the citation counts of non-arXiv papers with realisations of the same number of randomly selected telbib papers from the same year. For the years of our study, the non-arXiv sets show particularly high fractions of papers with low citations (0–10), followed by only a few papers with higher citations (> 30 citations for publication years 2010–2012; > 20 for publication years 2013–2014; and > 10 for publication years 2015–2017). Two examples from publication years 2010 and 2013 are shown in Figure 3.

### Conclusion

Using the ESO Telescope Bibliography (telbib) as a testbed, we investigated the fraction of refereed papers published between January 2010 and July 2017 that were submitted to arXiv/astro-ph and are therefore free to read. Our study revealed an increasing fraction of papers posted on arXiv, from 88.5 % in 2010 to 96.3 % in 2016. The percentage for 2017 (94.3 %) should be treated with caution as the year is not yet complete and the fraction may change. Why some papers are not posted on arXiv is unclear, as there are no significant trends among this group in terms of ESO facilities used, author affiliations or journals involved. A survey among these authors suggested various motivations for non-submission, with the top reason being a lack of time for the submission process.

A comparison of the average citation counts of arXiv *vs.* non-arXiv papers revealed that on average, papers published between 2010 and 2016 that were made available on the e-print server receive a factor of 2.8 more citations. Simulations suggested that these findings are statistically significant. While the differences in citations seem striking at first glance, we note that the sample of non-arXiv papers investigated is small (on average, 55 papers per year). Posting manuscripts on arXiv certainly enhances their visibility among the community, but we cannot draw a definitive conclusion as to whether these differences in citations are solely caused by submission to arXiv or whether other factors are at play.

Based on this study of telbib papers, we conclude that an almost complete availability of the core literature in astronomy through green OA puts the community in an advantageous situation regarding both literature access and dissemination.

#### Acknowledgements

We thank Wolfgang Kerzendorf and Jason Spyromilio for inspiring discussions (and a constant supply of chocolate). This research has made use of NASA's Astrophysics Data System; thank you so much ADS team for your excellent service to the community!

#### Links

[1] telbib database: telbib.eso.org
[2] telbib methodology: www.eso.org/sci/libraries/telbib_methodology.html
[3] ESO Libraries publications: www.eso.org/sci/libraries/useful_links/publications.html
[4] SPARC Open Access: https://sparcopen.org/open-access
[5] Creative Commons licensing: https://creativecommons.org/share-your-work/licensing-types-examples
[6] arXiv submission rate statistics: https://arxiv.org/help/stats/2016_by_area/index

Report on the ESO and Excellence Cluster Universe Workshop

# Galaxy Ecosystem: Flow of Baryons through Galaxies

held at ESO Headquarters, Garching, Germany, 24 –28 July 2017

Vincenzo Mainieri[1]
Paola Popesso[2]

[1] ESO
[2] Technical University of Munich, Germany

This conference focussed on the "baryon cycle", namely the flow of baryons through galaxies. The following aspects were discussed: a) the gas inflow into systems through streams of pristine gas or as drizzles of recycled material; b) the conversion of this gas into stars; and c) the ejection of gas enriched with heavy elements through powerful outflows. Understanding these different but mutually connected phases is of fundamental importance when studying the details of galaxy formation and evolution through cosmic time. This conference was held following the month-long workshop of the Munich Institute for Astro- and Particle Physics (MIAPP)[1] entitled: "In & out: What rules the galaxy baryon cycle?" It therefore provided an opportunity to share the main outcomes of the MIAPP workshop with a larger audience, including many young outstanding scientists who could not attend the MIAPP workshop.

In total, the conference attracted 132 participants who attended 67 talks over the